%% file: main.tex
\documentclass[manuscript]{acmart}
\setcopyright{none}
\usepackage{algorithmic}
\usepackage{graphicx}
\usepackage{textcomp}
\usepackage[table]{xcolor}
\usepackage{multirow}
\usepackage{algorithm}
\usepackage{subfigure}
\usepackage{listings}
\usepackage{url}
\usepackage{hyperref}

\AtBeginDocument{%
  }

\begin{document}

\title{Isolating Compiler Bugs through Compilation Steps Analysis}

\newcommand{\PKUinstitution}{Key Laboratory of High Confidence Software Technologies (Peking University), Ministry of Education; School of Computer Science, Peking University}

\author{Yujie Liu}
\orcid{0009-0001-1973-2792}
\affiliation{%
  \institution{\PKUinstitution}
  \city{Beijing}
  \country{China}}
\email{douya@stu.pku.edu.cn}

\author{Mingxuan Zhu}
\affiliation{%
  \institution{\PKUinstitution}
  \city{Beijing}
  \country{China}}
\email{zhumingxuan@stu.pku.edu.cn}

\author{Shengyu Cheng}
\affiliation{%
  \institution{ZTE Corporation}
  \city{Beijing}
  \country{China}}
\email{cheng.shengyu@zte.com.cn}

\author{Dan Hao}
\authornote{Corresponding author}
\affiliation{%
  \institution{\PKUinstitution}
  \city{Beijing}
  \country{China}}
\email{haodan@pku.edu.cn}


\renewcommand{\shortauthors}{Liu et al.}

\newcommand{\myName}{CompSCAN}
\newcommand{\todo}[1]{\textcolor{red}{[TODO] #1}}

\begin{abstract}
Compilers are essential to software systems, and their bugs can propagate to dependent software. Ensuring compiler correctness is critical. However, isolating compiler bugs remains challenging due to the internal complexity of compiler execution. Existing techniques primarily mutate compilation inputs to generate passing and failing tests, but often lack causal analysis of internal steps, limiting their effectiveness.

To address this limitation, we propose \myName, a novel compiler bug isolation technique that applies analysis over the sequence of compilation steps. \myName\ follows a three-stage process: (1) extracting the array of compilation steps that leads to the original failure, (2) identifying bug-causing steps and collecting corresponding compiler code elements, and (3) calculating suspicious scores for each code element and outputting a suspicious ranking list as the bug isolation result. 

We evaluate \myName\ on 185 real-world LLVM and GCC bugs. 
Results show that \myName\ outperforms state-of-the-art techniques in both effectiveness and efficiency. \myName\ successfully isolates 50, 85, 100, and 123 bugs within the Top-1/3/5/10 ranks, respectively.
Compared with ETEM and ODFL, two state-of-the-art compiler bug isolation techniques, \myName\ achieves relative improvements of 44.51\% / 50.18\% / 36.24\% / 24.49\% over ETEM, and 31.58\% / 49.12\% / 44.93\% / 21.78\% over ODFL on those metrics. Moreover, \myName\ runs faster on average per bug than both baselines.
\end{abstract}




\maketitle

\input{CompSCAN/1-introduction}

\input{CompSCAN/2-motivation}
\input{CompSCAN/3-approach}
\input{CompSCAN/4-experiment_setup}
\input{CompSCAN/5-results_and_analysis}

\input{CompSCAN/6-discussion}
\input{CompSCAN/7-related_work}
\input{CompSCAN/8-conclusion}

\newpage

\bibliographystyle{ACM-Reference-Format}
\bibliography{my_ref.bib}

\end{document}

%% file: CompSCAN/1-introduction.tex
\section{Introduction}\label{sec:introduction}

Compilers are fundamental software systems that translate programs from one language into another, such as from a high-level language to a machine-executable language. As a large amount of software depends on compilers to execute correctly, ensuring compiler correctness is critical. However, compiler debugging is time-consuming and labor-intensive~\cite{sun2016survey-compiler-bugs}, posing a significant challenge to compiler developers. Automated techniques for compiler debugging are therefore of high importance. Compiler bug isolation plays a central role in compiler debugging: given a buggy compiler and a triggering test input, the goal of compiler bug isolation is to isolate the elements of the compiler’s source code responsible for the bug.

Due to the inherent complexity and large code bases of modern compilers, general fault localization techniques, such as spectrum-based fault localization~\cite{SBFL-Ochiai, SBFL-tarantula} and mutation-based fault localization~\cite{MUSE, Metallaxis}, often perform poorly when directly applied to compiler bugs, both in terms of accuracy and efficiency~\cite{DiWi, RecBi}.

To address this, prior work has proposed several techniques to assist compiler bug isolation. 
For example, Regehr et al. introduced C-Reduce~\cite{C-Reduce}, and Wang et al. proposed DuoReduce~\cite{wang2025duoreduce}. These techniques leverage delta debugging for compiler test case reduction, which can assist compiler bug isolation by simplifying failure-inducing inputs. However, they do not directly pinpoint the suspicious compiler program elements.
Beyond test case reduction, other research areas are also related to compiler bug isolation. For instance, compiler fuzzing techniques (e.g., Csmith~\cite{Csmith}, EMI~\cite{EMI}) automatically generate diverse test programs to expose compiler failures, while differential testing frameworks (e.g., Alive2~\cite{Alive2}) verify semantic equivalence across compiler transformations.
These methods help discover or reproduce compiler bugs but, like test case reduction, they do not directly localize suspicious code within the compiler itself.
 
Other research has proposed techniques that isolate suspicious program elements directly, which have shown some effectiveness. 
DiWi~\cite{DiWi} mutates test programs to create passing variants (witness programs) and compares their coverage to identify suspicious code elements. RecBi~\cite{RecBi} adds more mutation strategies and uses reinforcement learning~\cite{reinforcement-learning, RL-ZTE-1} for better mutations. ETEM~\cite{ETEM} further improves this with feature mutations and joint RL networks. ODFL~\cite{ODFL} focuses on optimization bugs by exploring passing/failing fine-grained optimization settings. 

Existing techniques treat the compiler as a black box and isolate compiler buggy elements by modifying the input of the compiler (e.g., compiler test cases/optimization options) and analyzing the differences between passing and failing compiler inputs, which limits their ability to pinpoint the compiler elements that cause the bug directly. 
Lacking internal insight into the compilation process, these techniques rely on large test sets and statistical correlations to infer bug-relevant code, resulting in bug isolation with limited precision and efficiency.
For example, ETEM takes about an hour per bug to generate enough test cases for the coverage statistic analysis and needs to run five times to reduce randomness in test generation. 
These limitations highlight the need for directly reasoning about the relationship between the compilation process and the observed failures.

To overcome this limitation of existing techniques, in this paper, we propose \myName~(\textbf{Comp}ilation \textbf{S}tep \textbf{C}ausal \textbf{AN}alysis), a compiler bug isolation technique that improves both effectiveness and efficiency through precise analysis of compilation steps, rather than modifying compiler inputs.

\begin{figure}
    \centering
    \includegraphics[width=0.9\linewidth]{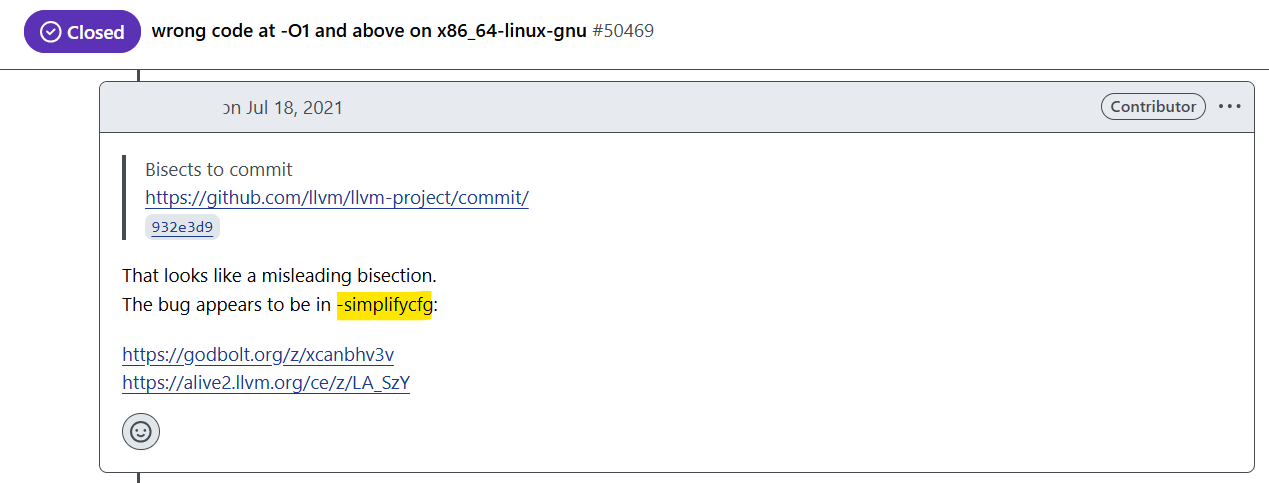}
    \caption{Bug Report For LLVM Bug \#50469}
    \label{fig:50469-bug-report}
    \Description{A screenshot of the GitHub comment thread for LLVM Bug \#50469, where a developer notes that the bug appears to be in the -simplifycfg pass.}
\end{figure}

Our key insight is that the compilation process can be naturally divided into discrete steps, and we can improve the efficiency of compiler bug isolation by directly identifying the compilation steps associated with a bug. Common compilers can typically be divided into multiple components, coarsely as frontend, middle-end, and backend, and further refined if needed. 
Modern compilers often support selective execution of individual components. For instance, GCC provides flags to enable or disable specific optimizations~\cite{GCC, GCC-optimization-flags}, while LLVM~\cite{LLVM} and the MLIR compiler~\cite{MLIR} allow users to choose which passes to execute~\cite{LLVM-Passes, MLIR-Passes}. 

By identifying the compilation steps that lead to a failure and isolating the corresponding source code, we can achieve more accurate and efficient bug isolation. Our observation of GCC and LLVM bug reports indicates that developers often begin debugging by reasoning about which compilation phases may have caused the observed misbehavior. Figure~\ref{fig:50469-bug-report} illustrates an example from LLVM bug report \#50469: after the bug was reported, the first response from the developer highlights that the pass \texttt{simplifycfg} was identified as the source of the failure. This observation underscores the practical relevance of causal reasoning. \myName\ automates this reasoning over compilation steps in general-purpose compilers, bridging the gap between human intuition and automated bug isolation.

However, designing \myName\ poses two main challenges. 
First, identifying the bug-inducing steps is non-trivial, as failures may originate from earlier steps due to complex interdependencies within the compilation process. 
Second, compilation steps can involve a large portion of the compiler's source code. Accurately pinpointing the most suspicious compiler code elements associated with a bug-inducing step, therefore, requires an efficient method for computing suspiciousness.

Existing techniques fall short in addressing these challenges.
Delta debugging–based techniques  (e.g., C-Reduce, DuoReduce) aim to minimize failing test cases rather than analyze the compiler’s internal behavior. Consequently, they cannot reveal which specific compilation steps or source components are responsible for a failure.
In contrast, existing compiler bug isolation techniques (e.g., ETEM, ODFL) rely heavily on large-scale program mutations and statistical correlations between passing and failing executions. Their accuracy degrades when failures propagate through interdependent compilation steps or when each step involves a significant portion of the compiler’s codebase.

To address these challenges, \myName\ performs step-driven analysis over compilation processes in three stages. First, it extracts the compilation process of a buggy compiler and a failing test case as an \textit{Initial Failing Sequence}, representing each distinct step in execution order. Second, \myName\ identifies \textit{bug-causing steps}, which with direct or indirect causal influence on the observed failure, and collects the corresponding compiler code elements. To reduce noise from unrelated code, a \emph{tail step pruning} strategy is applied to minimize downstream effects of each step. 
Finally, \myName\ computes suspiciousness scores for all affected compiler elements based on coverage differences observed when removing each bug-causing step. The resulting scores are aggregated and ranked to highlight the source elements most likely responsible for the failure.

We evaluate \myName\ on a dataset of 185 real-world compiler bugs, consisting of all 155 bugs from prior studies~\cite{ODFL, ETEM} and 30 newly collected LLVM bugs. The new LLVM bugs were gathered from the official LLVM GitHub repository following the same collection protocol used in prior work(see Section~\ref{sec:experiment_setup}), ensuring consistency while reflecting more recent and realistic compiler failures. This large-scale evaluation, spanning both LLVM and GCC, enables a fair and representative comparison.

Experimental results demonstrate that \myName\ achieves higher accuracy and faster runtime compared to state-of-the-art techniques ETEM and ODFL. On a large-scale set of 125 LLVM bugs, \myName\ achieves Top-1/3/5/10 accuracy of 32, 53, 62, and 78 bugs, respectively, corresponding to improvements of 75.82\%, 105.43\%, 72.22\%, and 40.79\% over ETEM, and 33.33\%, 51.43\%, 51.22\%, and 32.20\% over ODFL. \myName\ also isolates each bug in 168.16 seconds on average, much faster than ETEM (5×3600s) and ODFL (523.83s). These results highlight that \myName\ can effectively isolate compiler bugs with higher precision while maintaining efficient runtime. 
To demonstrate generalizability, we further evaluate \myName\ on 60 GCC bugs. Despite differences in compiler architecture and compilation pipelines, \myName\ shows consistent improvements over existing techniques, confirming that our step-based analysis is effective across different compiler infrastructures.

This paper makes the following key contributions:
\begin{itemize}
\item \textbf{Approach.} We present \myName, a novel approach that leverages the structure of the compilation process to isolate compiler bugs. By analyzing individual compilation steps, \myName\ identifies the steps responsible for a failure and ranks the associated compiler code elements by suspiciousness, enabling more precise and efficient bug isolation than traditional techniques.

\item \textbf{Tool.} We implement \myName\ as a practical compiler bug isolation tool based on Python and the Gcov instrumentation tool.
\item \textbf{Evaluation.} We extend the dataset used in previous studies to include 185 bugs and conduct extensive experiments based on this dataset. Results show that \myName\ consistently outperforms all state-of-the-art techniques in both effectiveness and runtime.
\end{itemize}

The rest of this paper is organized as follows. Section~\ref{sec:motivation} explains the motivation. Section~\ref{sec:approach} describes our approach design. Section~\ref{sec:experiment_setup} details the experiment setup. Section~\ref{sec:results_analysis} presents and analyzes the experimental results. Section~\ref{sec:discussion} discusses validity threats. Section~\ref{sec:related_work} reviews related work. Section~\ref{sec:conclusion} concludes the paper.


%% file: CompSCAN/2-motivation.tex
\section{Motivation}\label{sec:motivation}

To motivate our approach, we present some real LLVM bugs to show the limitations of existing bug isolation techniques. We then show how step-level compilation causal analysis enables more precise and efficient isolation, and outline the key challenges involved.

\begin{lstlisting}[
  language=bash,
  caption={LLVM Bug \#50469},
  basicstyle=\ttfamily\footnotesize,
  columns=fullflexible,
  backgroundcolor=\color{gray!5},
  breaklines=true,
  showstringspaces=false,
  commentstyle=\color{gray},
  keywordstyle=\color{black},
  xleftmargin=1em,
  framexleftmargin=1em,
  label={lis:50469-test}
]
$ clang -O0 fail.c; ./a.out
1
$ clang -O1 fail.c; ./a.out
-1
\end{lstlisting}

\begin{figure}[htbp]
\centerline{\includegraphics[width=0.45\textwidth]{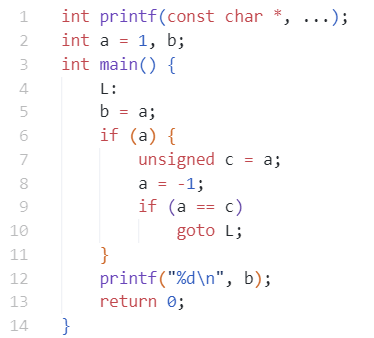}}
\caption{Input Program of \#50469}
\label{fig:50469-c}
\Description{A C source program that causes LLVM internal error \#50469 under certain optimization flags.}
\end{figure}

\begin{figure}[htbp]
\centerline{\includegraphics[width=0.9\textwidth]{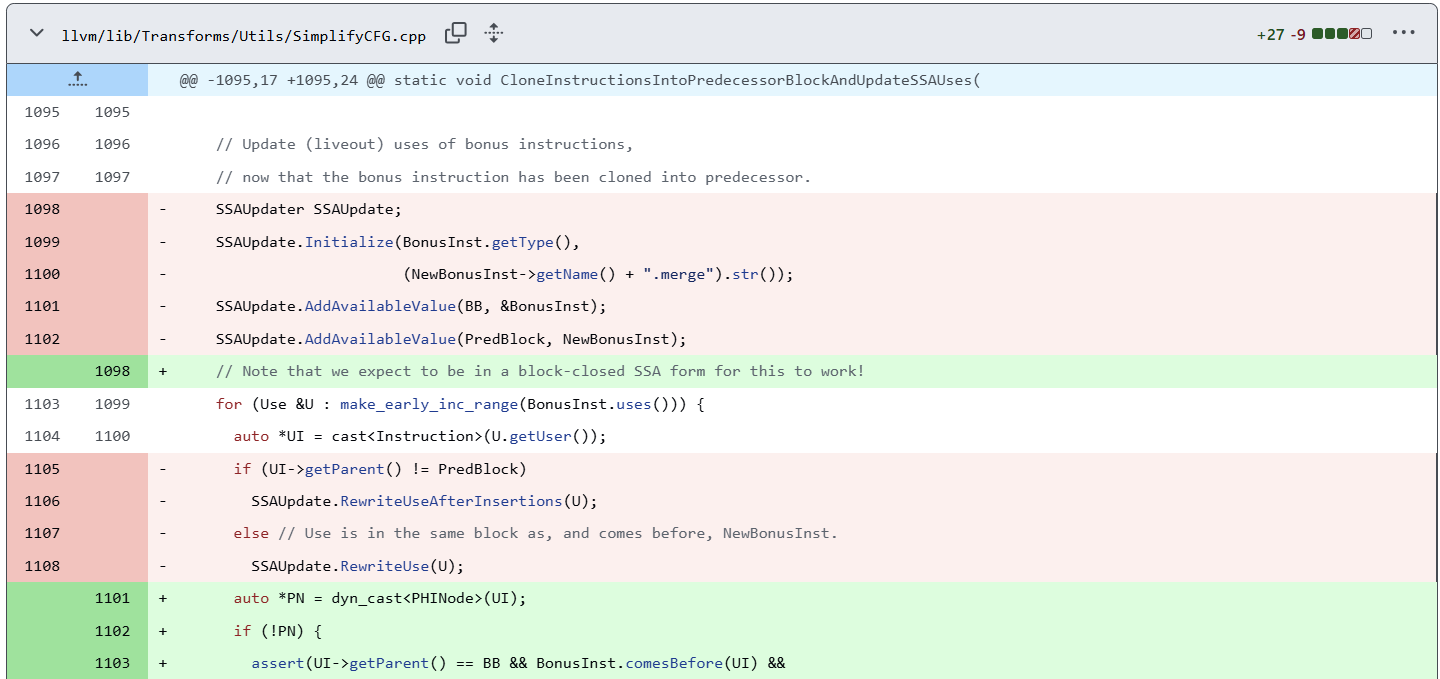}}
\caption{Fix Commit of \#50469}
\label{fig:50469-root}
\Description{A screenshot of the GitHub diff view for the fix commit of LLVM Bug \#50469, showing code changes made to the SimplifyCFG.cpp file.}
\end{figure}

\subsection{Limitations of existing techniques}

Listing~\ref{lis:50469-test} presents a real-world bug in LLVM (Bug \#50469)~\cite{llvm-bug-50469}. When compiling the input program shown in Figure~\ref{fig:50469-c} with the optimization flags \texttt{-O0} and \texttt{-O1}, the resulting executables (i.e., \texttt{./a.out}) produce different outputs. Since compiler optimizations should preserve the program's semantics, this discrepancy indicates a compiler bug.

Figure~\ref{fig:50469-root} shows the key part of the fix commit for this bug. The root cause lies in the \texttt{SimplifyCFG} pass\footnote{In LLVM, a pass is a compiler optimization component that traverses parts of a program to either gather information or apply transformations~\cite{LLVM-Passes}.}: the buggy compiler updates the phi in the successor with a value from the new/updated block, but that's wrong for a load of memory that might have been overwritten.

When applying existing bug isolation techniques to this case, witness-based techniques, such as ETEM, attempt to generate new test programs by mutating the input program (i.e., \ref{fig:50469-c}), while optimization-search-based techniques, such as ODFL, explore different optimization configurations (e.g., beyond just \texttt{-O0} or \texttt{-O1}) and apply these configurations to compile the original input program to generate new test cases.
These techniques then apply coverage-based statistical analysis to rank suspicious compiler elements. 

However, the root cause file, which is named \texttt{llvm/lib/Transforms/Utils/SimplifyCFG.cpp}, is ranked only 45th by ETEM and 11th by ODFL. The limited precision results from a lack of analysis of the internal compilation process. 

ETEM, for example, relies on source-level mutations. However, the SimplifyCFG pass (i.e., the buggy component of the LLVM compiler implemented in \texttt{SimplifyCFG.cpp}) operates on LLVM IR. The source-to-IR mapping is often imprecise (one line of source code may correspond to multiple IR instructions or none at all).
As a result, ETEM’s mutations fail to meaningfully affect the buggy IR-level behavior, limiting its effectiveness in isolating the root cause.

ODFL searches for compiler optimization configurations that can change the outcome of a test from failing to passing or vice versa. However, each optimization flag influences a broad range of compiler behaviors, making it difficult to precisely pinpoint the root cause within LLVM's complex optimization pipeline. Moreover, high-quality configurations that consistently trigger a pass or fail are scarce. For instance, in the case of LLVM bug \#50469, ODFL only discovered six additional compilations, each containing numerous irrelevant compiler options. Notably, even the minimal configuration that triggered the bug included 34 additional options besides \texttt{SimplifyCFG}, which limits the efficiency of bug isolation.

This example illustrates the limitation of treating compilers as a black box and motivates the need for causality-driven analysis of the compilation process used in \myName.

\subsection{Causal-driven analysis over compilation steps}

Building on the above example, we now explore how analyzing the compilation process at the step level can lead to more effective bug isolation. Modern compilers (such as LLVM in this example) allow the compilation pipeline to be broken down into individually executable or skippable steps, such as frontend processing, middle-end transformation and analysis, and backend code generation and optimizations that can be enabled or disabled separately. In the case of LLVM Bug \#50469, the compilation can be decomposed into 107 distinct steps (including middle-end passes such as \texttt{SimplifyCFG}), each step corresponding to an independently controllable action.

Our key insight is that by isolating the compilation steps that have a causal relationship with the final failing test and identifying the compiler source code statements associated with these steps, compiler bug isolation can be performed more accurately and efficiently. In practice, many developers adopt a similar strategy when debugging: they first determine which compilation step is responsible for the observed failure, as illustrated in Figure~\ref{fig:50469-bug-report}.

However, labeling the bug-causing steps and mapping them to relevant compiler source code presents two challenges.

\begin{figure}
\centering
\includegraphics[width=0.9\linewidth]{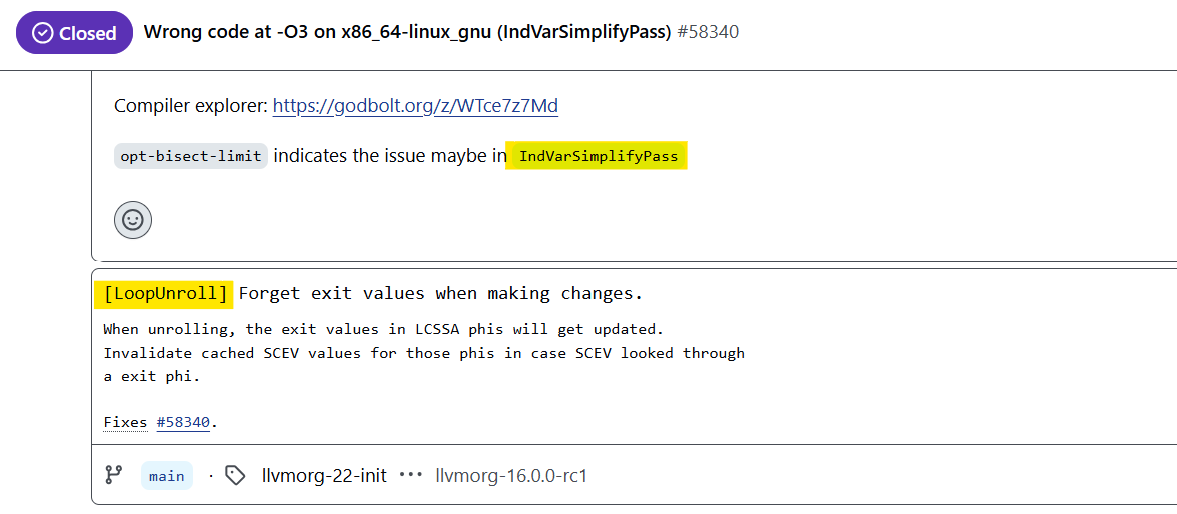}
\caption{Bug Report and Fix Commit Message for LLVM Bug \#58340}
\label{fig:58340-bug-report}
\Description{A screenshot showing both the GitHub issue and fix commit for LLVM Bug \#58340. The issue discussion indicates that the first failing pass is IndVarSimplifyPass, while the fix commit shows that the actual faulty pass is LoopUnroll.}
\end{figure}

\textbf{Challenge 1: Identifying Buggy Steps.} 
Faulty behavior does not always indicate the faulty step. Figure~\ref{fig:58340-bug-report} shows an example for such cases, in LLVM Bug \#58340~\cite{llvm-bug-58340}, the output after the \texttt{IndVarSimplify} pass first exhibits faulty behavior, but the root cause lies in the earlier \texttt{LoopUnroll} pass. Specifically, \texttt{LoopUnroll} modifies the control flow graph and updates PHI nodes in loop exit blocks but fails to invalidate outdated analysis results cached by \texttt{ScalarEvolution} (SCEV). As a result, \texttt{IndVarSimplify}, which relies on SCEV, consumes stale and incorrect information, leading to faulty transformations. Although the output of \texttt{LoopUnroll} itself appears correct, it silently introduces an incorrect internal state that manifests later in the pipeline. \myName\ temporarily skipping individual steps to address this challenge: for each compilation step, if omitting this step causes the failure to disappear, we label it as a \textit{bug-causing step}. In the case of Bug \#58340, skipping \texttt{LoopUnroll} eliminates the failure, revealing it as causally responsible for the bug.
Note that traditional Delta Debugging and its extensions, such as C-Reduce and DuoReduce, cannot address this challenge because they focus on minimizing failing test cases rather than reasoning about causality between compilation steps. In particular, they do not systematically test the effect of omitting individual steps, and thus cannot detect cases where an earlier step silently corrupts the state that triggers failures in later steps.

\textbf{Challenge 2: Mapping Relevant Compiler Source Code.} 

The goal for bug isolation is to isolate the specific compiler source code responsible for the bug. However, due to the compiler’s large code base, mapping the bug-causing steps to the corresponding compiler source code is nontrivial. 
The compiler’s large code base introduces substantial noise when attempting to associate individual steps with the relevant source-level statements. 
As a result, directly pinpointing the buggy code associated with the bug-causing steps remains challenging.

Returning to the first example, LLVM Bug \#50469, \myName\ identifies three steps responsible for the bug: the \texttt{InstCombine} pass, the \texttt{LICM} pass, and the \texttt{SimplifyCFG} pass. However, these steps still encompass a large portion of the compiler source code. Even a minimal execution coverage that includes only these three steps and reproduces the bug contains 20,168 lines of code across 720 distinct files, making it extremely challenging to isolate the files most likely responsible for the bug.

To address this challenge, \myName\ applies \emph{tail step pruning}, a technique that identifies and removes coverage differences for each bug-causing step in reverse order while minimizing the number of statements by discarding steps whose removal does not affect the observed failure. In this example, the number of potentially relevant compiler statements is reduced from 20,168 to just 2,029 (coverage difference after removing \texttt{InstCombine}), 1,472 (after removing \texttt{LICM}), and 408 (after removing \texttt{SimplifyCFG}), substantially narrowing the scope for further analysis. 
Moreover, \myName\ employs a suspiciousness computation based on coverage differences: for each bug-causing step, statements in its coverage difference are assigned suspiciousness scores inversely proportional to the size of the difference, and these scores are then aggregated at the file level to complete compiler bug isolation.
As a result, \myName\ ranks the root-cause file 3rd, dramatically outperforming prior techniques, which ranked it 45th and 11th.

Based on these observations, we design \myName\ through causal analysis of compilation steps. The detailed design of our approach is presented in Section~\ref{sec:approach}.

%% file: CompSCAN/3-approach.tex
\section{Approach}\label{sec:approach}

\begin{figure*}[htbp]
\centerline{\includegraphics[width=\textwidth]{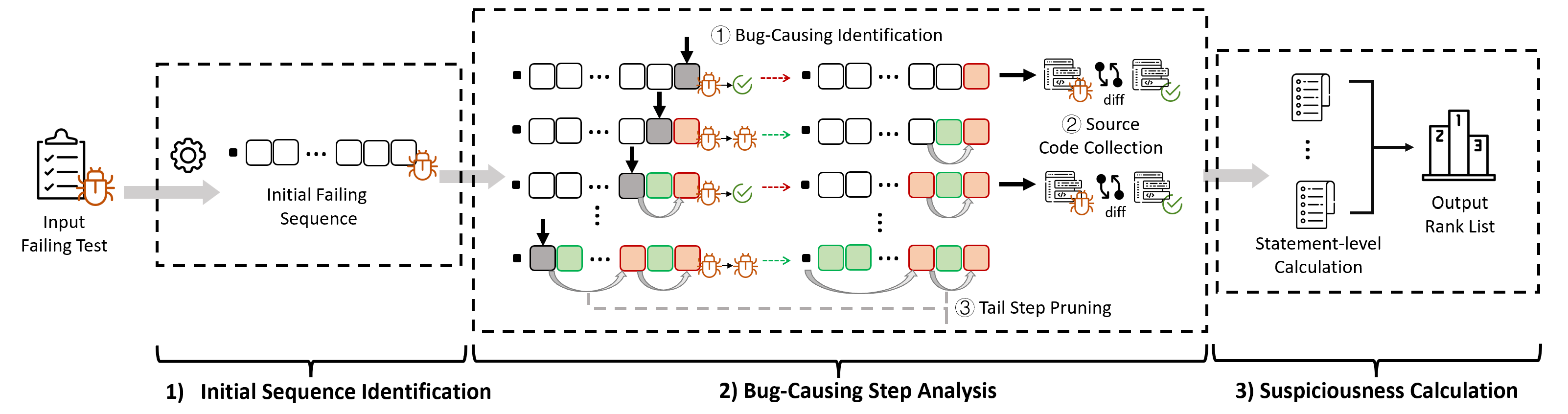}}
\caption{Overview of \myName}
\label{fig:approach-overview}
\Description{Overview of the proposed approach CompSCAN. Given a failing test as input, the system first performs Initial Failing Sequence Identification to obtain the sequence of executed compilation steps. Then, during Bug-Causing Step Analysis, each step in the sequence is iteratively removed in reverse order to observe the change in test outcome and collect execution coverage. Finally, the Suspiciousness Calculation computes the suspiciousness ranking of compiler statements based on coverage differences.}
\end{figure*}

In this paper, we propose a novel compiler bug isolation technique named \myName. While prior techniques rely on generating diverse inputs, whether through test program mutations or optimization configuration searches, to isolate compiler bugs, \myName\ introduces a fundamentally new approach by causally analyzing the internal sequence of compilation process steps and their corresponding compiler source code, enabling a more direct reasoning about the final bug appearance. Figure~\ref{fig:approach-overview} presents an overview of our approach, which consists of the following three key stages:

\begin{itemize} 
    \item \textbf{Initial Failing Sequence Identification}: 
    Given a failing compilation process, we record the full sequence of executed compilation steps as a step array, which serves as the starting point for further analysis.
    \item \textbf{Bug-Causing Step Analysis}: 
    \myName\ analyzes the step array by trying to temporarily skip each step and see if the bug still occurs. Steps whose omission prevents the failure are identified as bug-causing steps. \myName\ then maps the identified bug-causing steps to the corresponding compiler code, and applies tail-pruning to reduce irrelevant statements.
    \item \textbf{Suspiciousness Calculation}:  
    Finally, \myName\ computes a suspiciousness score for each compiler code element based on the cardinality of the code-element set associated with each bug-causing step and produces a ranked list of suspicious elements as the bug-isolation result.
\end{itemize}

Compared to existing input-modification-based techniques, \myName\ offers a different, compile-step-driven view of the buggy compilation process, enabling more precise and efficient compiler bug isolation.

The following subsections detail each stage: Section~\ref{sec:approach:initial-sequence} covers Initial Failing Sequence Identification; \ref{sec:approach:bug-causing-analysis} describes Bug-Causing Step Analysis; and \ref{sec:suspicious-calculation} explains Suspicious Calculation.

\subsection{Initial Failing Sequence Identification}\label{sec:approach:initial-sequence}

To enable stepwise analysis of the compilation process, we first identify the sequence of compilation steps responsible for triggering the observed failure.

The compilation process of a program typically proceeds through a series of phases. 
For example, at a coarse-grained level, the compilation process is often divided into three major stages: the frontend, middle-end, and backend. The frontend parses and translates the source code into an intermediate representation (IR); the middle-end performs a sequence of analyses and optimizations; and the backend generates target machine code, often applying further optimizations.
In practice, each of these stages is further divided into a sequence of fine-grained transformations or analysis fragments (e.g., pass in LLVM and MLIR~\cite{LLVM-Passes, MLIR-Passes}, optimizations in GCC~\cite{GCC-optimization-flags}). Each fragment incrementally transforms the program based on the output of the previous one, forming a natural sequence of compilation steps. Each such step corresponds to a distinct part of the compiler's codebase, often much smaller in scope compared to the full compiler. 

Our insight is that if we can isolate each compilation step such that it can be independently executed or selectively skipped, we can determine its impact on the final compilation result. This enables more targeted and efficient compiler bug isolation by narrowing the analysis to only those compiler components directly responsible for the failure.
Therefore, in \myName, we model the compilation process as an array of discrete steps, where each step corresponds to a logical fragment of the compilation process that can be individually executed or skipped. For instance, many compilers allow selectively applying or disabling individual middle-end or backend analyses and optimizations (e.g., using \texttt{opt --passes="..."} in LLVM or fine-grained optimization flags in GCC). Each such configurable transformation is treated as a compilation step. By structuring the compilation this way, we enable systematic reasoning about the impact of each step on the final buggy outcome.

Given a buggy compiler and a failing input (i.e., a test program together with the compilation options that trigger the bug), \myName\ first executes the entire compilation process and records all intermediate steps that can be individually enabled or disabled. These steps typically correspond to configurable components accessible through debugging interfaces (e.g., \texttt{--print-pipeline-passes} in LLVM or \texttt{-Q --help=optimizers} in GCC).
This procedure yields a stepwise decomposition of the compilation process, represented as an array we termed the \textit{Initial Failing Sequence}: an ordered collection of compilation steps that, when executed in sequence, reproduce the observed failure. The Initial Failing Sequence provides the foundation for subsequent analysis, wherein \myName\ evaluates the contribution of each step to the overall outcome and pinpoints those responsible for triggering the failure.




\subsection{Bug-Causing Step Analysis}\label{sec:approach:bug-causing-analysis}

After constructing the Initial Failing Sequence, \myName\ proceeds to isolate the faulty component within the compiler by determining which steps in the sequence are responsible for triggering the bug and analyzing their corresponding source code, directly pinpointing the root cause of the failure.
This analysis consists of three components: Section~\ref{sec:bug-causing-identification} introduces the identification of bug-causing steps; Section~\ref{sec:code-collection} details the collection of the corresponding compiler source code; and Section~\ref{sec:tail-pruning} describes the reduction of irrelevant code originating from unrelated steps.

\subsubsection{Bug-Causing Step Identification}\label{sec:bug-causing-identification}

We define steps in the Initial Failing Sequence that are responsible for triggering the bug as \textit{Bug-Causing Steps}. Specifically, a step is considered a Bug-Causing Step if its removal causes the failure observed in the final test outcome to disappear. This definition is grounded in a principle commonly used in delta debugging~\cite{zeller-delta-debugging}: given a set of changes $\{c_1, c_2, \dots, c_n\}$ applied to a buggy program, a change $c_i$ is irrelevant to the bug if its presence does not affect the manifestation of the failure. In our context, the removal of a compilation step can be interpreted as such a change; therefore, a step is deemed potentially causal (i.e., suspicious) if omitting it eliminates the bug. Identifying these steps is instrumental for isolating the root cause of the failure.

To identify all bug-causing steps, \myName\ traverses the Initial Failing Sequence in reverse order (see~\ref{sec:tail-pruning} for the rationale) and temporarily skips each step. For each step under consideration, we construct a modified compilation sequence by omitting that step and executing the remainder. If the omission eliminates the failure (i.e., the original test no longer fails), the skipped step is classified as bug-causing; otherwise, it is deemed non-essential to the failure.
Unlike Delta Debugging, which aims to find a minimal subset of changes or inputs that still reproduces the failure, our technique explicitly identifies each step that is causally responsible for the bug. Delta Debugging may produce a reduced failing sequence, but not all steps in that sequence are necessarily bug-causing. In contrast, \myName\ provides a step-level causal analysis, pinpointing the precise compilation steps whose removal eliminates the failure.

It is important to note that a Bug-Causing Step does not necessarily correspond to the location of the actual fault. Some steps may only provide the necessary preconditions for a subsequent buggy step to manifest. 
Nonetheless, identifying these steps remains valuable, as the downstream compiler code affected by them still contains the actual faulty elements. In subsequent stages, \myName\ collects these relevant compiler code elements and assigns them appropriate suspiciousness scores.

\subsubsection{Source Code Collection}\label{sec:code-collection}

After identifying the bug-causing steps, we aim to determine the corresponding compiler source code responsible for the failure to achieve finer-grained bug isolation.
A natural idea is to execute each bug-causing step independently, using the IR produced by the preceding step as input, and then collect its execution coverage. However, this strategy has severe practical limitations. Executing a compilation step in isolation may break internal dependencies between steps, such as shared analysis results or runtime states, leading to incorrect behavior and inaccurate coverage data. As the example in Section~\ref{sec:motivation}, LLVM's \texttt{LoopUnroll} pass influences \texttt{ScalarEvolution} (SCEV) analysis, which may later be reused by \texttt{IndVarSimplify} pass to simplify induction variables. Executing a step in isolation can break such dependencies, resulting in altered behavior.

To address this challenge, we adopt a reverse analysis strategy. Rather than executing each step in isolation, we selectively execute the complete compilation sequence while temporarily skipping each bug-causing step to collect coverage differences. For each bug-causing step, we obtain coverage data from both the original failing execution (with all steps executed) and the modified execution (with the step omitted). By comparing these coverages, we can infer the compiler source code associated with the bug-causing step, without violating internal dependencies between steps.

It is important to note that, through this definition of bug-causing steps and the associated code collection procedure, \myName\ can handle complex scenarios such as \#58340 illustrated in Section~\ref{sec:motivation}, where an intermediate step produces an erroneous program state that subsequently leads to a faulty output in another step. In this case, removing the step that generates the incorrect state causes the failure to disappear; consequently, this step is identified as bug-causing, and the relevant compiler code is collected. 
Similarly, \myName\ can handle cases where multiple steps jointly trigger a bug: the removal of any single step among them eliminates the failure, so all such steps are recognized as bug-causing, and the corresponding code for each step is collected.

\subsubsection{Tail Step Pruning}\label{sec:tail-pruning}

As described above, we collect coverage by selectively skipping each bug-causing step within the sequence. However, this mapping is complicated by the scale and complexity of modern compilers: they typically consist of large and intricate code bases, where a single step may affect numerous components across multiple modules. Consequently, identifying the precise source code associated with each step is highly nontrivial. 

In our context, this challenge implies that the coverage differences collected in Section~\ref{sec:code-collection} may contain a substantial amount of downstream changes of unrelated code execution. Such noise is undesirable, as it interferes with the isolation of the truly buggy code elements. Therefore, our goal is to ensure that, when collecting coverage differences for each bug-causing step, the number of irrelevant downstream statements is minimized.

To address this problem, we propose \textit{Tail Step Pruning}, a strategy that simplifies the step sequence after each bug-causing step by removing as many irrelevant steps as possible while maintaining efficiency. Tail Step Pruning integrates bug-causing step identification and code collection with a lightweight variant of delta debugging. 

Specifically, our objective is to collect the coverage differences of each bug-causing step (between execution and omission) while minimizing the number of subsequent steps affected. To achieve this, given the Initial Failing Sequence, \myName\ applies a divide-and-conquer strategy: \myName\  first splits the sequence into two halves and tests whether removing the latter half eliminates the failure. If the failure persists, the latter half is deemed irrelevant and discarded, and the first half is recursively examined. If the failure disappears, it indicates that at least one bug-causing step lies in the latter half, which is then further subdivided (e.g., testing the last quarter of the sequence). This process continues until only a single step remains. If removing that step eliminates the failure, it is identified as bug-causing, and the corresponding coverage differences are collected for subsequent bug isolation. At this point, all subsequent steps have already been examined, so all non-bug-causing steps have been removed to the greatest extent possible.

This strategy guarantees that, when collecting coverage differences for each bug-causing step, no irrelevant downstream steps are included, but only those that are causally related to the final failure remain. All unrelated steps are pruned efficiently during the process.

\subsection{Suspiciousness Calculation}\label{sec:suspicious-calculation}

\subsubsection{Statement-level Calculation}

After collecting the coverage information, \myName\ identifies suspicious compiler program elements by analyzing the coverage differences of bug-causing steps and assigning a suspiciousness score to each element.


\myName\ assigns suspiciousness scores to compiler source code statements to prioritize those most likely responsible for a bug. The key idea is simple: each \textit{bug-causing step} identified in the compilation sequence may cover multiple compiler statements, and the compiler statements that participate in the steps whose removal eliminates the failure are more suspicious. 

We formalize this idea by treating the effect of removing a bug-causing step as a \textit{mutant} $m$, and let $\text{diff}_m$ denote the set of compiler statements whose execution is affected by this step. Since each mutant (i.e., removing the step) may flip the failing test outcome (i.e., causing the original failing test to pass), we measure the flipping result of a mutant $m$ by the following formula.
\begin{equation}\label{eq:my-mut}
M(m) = 
\begin{cases}
1, & \text{if mutant m flips the failing test outcome;} \\
0, & \text{otherwise.}
\end{cases}
\end{equation}
This binary formulation clearly captures whether a mutant contributes observable evidence that helps identify the bug’s root cause.

The suspiciousness of a statement $s$ is then defined as:
\begin{equation}\label{eq:my-line}
S(s) = \max_{m \in \text{mut}(s)} M(m) \cdot \frac{1}{|\text{diff}_m|},
\end{equation}
where $\text{mut}(s)$ denotes the set of mutants that involve statement $s$. 
More precisely, a statement $s$ is considered \emph{involved} in a mutant $m$ if its execution status (i.e., whether it is executed) differs between the original failing run and the run after applying the mutation (i.e., removing the bug-causing step). 
Intuitively, statements involved in a mutant are potentially responsible for the bug, as their behavior changes when the failing outcome is flipped. 
In our setting, we identify these involved statements by computing the difference in compiler statement coverage between the original failing execution and the execution after removing the bug-causing step corresponding to mutant $m$, denoted as $\text{diff}_m$.
The normalization factor $1/|\text{diff}_m|$ captures the intuitive idea that when a mutant involves many statements, the suspiciousness assigned to each individual statement should be smaller. For example, a mutant involving all statements is less informative than one involving only a few.

This technique is inspired by mutation-based fault localization (MBFL)~\cite{MUSE, Metallaxis}, where each statement in a program is mutated and statements whose mutants change test outcomes are considered suspicious. In \myName, the removal of a bug-causing step is analogous to a mutation, but unlike traditional MBFL, a single step affects a large set of statements rather than one. Conceptually, the same principle applies: changes that flip a failing test outcome indicate higher suspiciousness, and each statement is scored based on its participation in such changes.

We can further provide a probabilistic justification for the factor $\frac{1}{|\text{diff}_m|}$ in Equation~\ref{eq:my-line}. 
Let $S$ be the total number of statements, $B$ the number of faulty statements, and $|\text{diff}_m|$ the size of mutant $m$'s coverage difference. Assuming a mutant flips the failing test only if it includes at least one faulty statement, and all statements are equally likely to appear in $\text{diff}_m$, Bayes' theorem gives:
\begin{equation}
P(s \text{ faulty} \mid m \text{ flips test}) = \frac{P(m \text{ flips test} \mid s \text{ faulty}) \cdot P(s \text{ faulty})}{P(m \text{ flips test})}.
\end{equation}
With $P(s \text{ faulty}) = B/|S|$ and $P(m \text{ flips test}) \approx |\text{diff}_m| \cdot B / |S|$ (valid when $B \ll |S|$ and $|\text{diff}_m| \ll |S|$), and assuming $P(m \text{ flips test} \mid s \text{ faulty}) \approx 1$, we have
\begin{equation}
P(s \text{ faulty} \mid m \text{ flips test}) \approx \frac{1}{|\text{diff}_m|}.
\end{equation}
Thus, the factor $\frac{1}{|\text{diff}_m|}$ naturally captures the inverse relationship between the size of the coverage difference and the likelihood that any individual statement is faulty, justifying the even distribution of mutant suspiciousness across $\text{diff}_m$.

\subsubsection{File-level Calculation}

Following prior compiler bug isolation work~\cite{DiWi, RecBi, ODFL, ETEM}, \myName\ performs file-level bug localization by default. To this end, we aggregate statement-level suspiciousness scores into file-level scores using the Rank-Sum (RS)~\cite{rank-survey} strategy following prior work~\cite{ETEM}. Specifically, statements in each file are ranked in descending order of suspiciousness, and a weighted average is computed to represent the file's overall suspiciousness. In RS, the weight $w_i$ for the $i$-th ranked statement among $n$ total statements is computed as:
\begin{equation}\label{eq:RS}
w_i = \frac{n+1-i}{\sum_{j=1}^n j}
\end{equation}

Therefore, the suspiciousness score of file $f$ is given by:
\begin{equation}
score(f) = w_i \cdot score(s_i) = \frac{2(n+1-i) \cdot score(s_i)}{n(n+1)}
\end{equation}
where $n$ is the number of statements in file $f$, and $score(s_i)$ is the suspiciousness score of the $i$-th ranked statement.

It is worth noting that \myName\ is not restricted to file-level bug isolation. By adjusting the aggregation level, \myName\ can be extended to support different granularities, such as function-level. We demonstrate this generality by also presenting function-level bug isolation results in Section~\ref{sec:results_analysis}.

%% file: CompSCAN/4-experiment_setup.tex
\section{Experiment Setup}\label{sec:experiment_setup}



In this study, we address the following research questions:
\begin{itemize}
    \item \textbf{RQ1}: How is the bug isolation effectiveness of \myName\ compared with existing techniques?
    \item \textbf{RQ2}: How does each component of \myName\ contribute to the final bug isolation result?
    \item \textbf{RQ3}: 
    To what extent is \myName\ generalizable? Can \myName\ be applied to compilers where step sequences are difficult to obtain?
\end{itemize}

\subsection{Benchmark}

For RQ1, we use the widely adopted open-source compiler LLVM~\cite{LLVM} as our experimental subject as its pass-based structure makes it relatively straightforward to identify compilation steps and their execution order. Moreover, LLVM has been the common target in prior compiler bug isolation studies~\cite{ETEM, ODFL}. In the experiment, we evaluate \myName\ on 125 real-world LLVM bugs: (1) 95 LLVM bugs are taken from existing datasets~\cite{ETEM, ODFL}, fully covering all the LLVM bugs used in prior work, and (2) another 30 bugs are newly collected from recent LLVM GitHub issues~\cite{llvm-github} to better assess the capability of each approach in isolating relatively newer bugs. To maintain consistency with previous studies and facilitate reproducibility, we adopt the same selection criteria as in prior work~\cite{DiWi, ETEM}, requiring that bugs (1) were fixed by LLVM developers, (2) include a failing test in the bug report, and (3) are reproducible in our experimental environment. 
To collect the new LLVM bugs, we manually examined issues on the LLVM GitHub repository in reverse chronological order, starting from the most recent issues available at the time of our experiments (March 2025). Whenever an issue satisfied the above criteria, it was recorded. Since this process involved substantial manual effort, we collected 30 new bugs.  
To investigate the generalizability of \myName\ (RQ3), we conduct experiments on both LLVM and GCC~\cite{GCC}, which together cover the two most frequently studied compilers in prior work on compiler bug isolation~\cite{DiWi, RecBi, ETEM, ODFL}. 
Unlike LLVM, GCC does not offer an option to directly expose the execution order of compilation passes, and some optimization steps partially overlap. As a result, \myName\ cannot fully exploit precise step sequencing on GCC, reflecting a limitation of our approach; nevertheless, it can still operate and provide meaningful bug isolation results, demonstrating the method’s generality.

In particular, in this experiment, we use 60 real-world GCC bugs, covering all GCC bugs that have been used in prior studies~\cite{DiWi, RecBi, ETEM, ODFL}. The GCC dataset is identical to that used in previous work.

Table~\ref{tab:dataset-info} summarizes the dataset, including the buggy LLVM/GCC version, total compiler source files, and files covered by the failing test.

\begin{table}[ht]
\small
\centering
\caption{Bug Dataset Information}
\begin{tabular}{cccc}
\hline
Compiler & Compiler Version & \#Files & \#CovFiles \\ 
\hline
LLVM & 3.9 - 20.0 & 81355 - 153745 & 895 - 1448 \\
GCC & 14.10.0 - 8.0.0 & 148889 - 191730 & 50 - 367\\
\hline
\end{tabular}
\begin{minipage}{\textwidth}
\vspace{1mm}
\#Files: The total number of files in the compiler project. \\
\#CovFiles: The number of files covered by the original failing test.
\end{minipage}
\label{tab:dataset-info} 
\end{table}

\subsection{Compared Approach}

For RQ1 and RQ3, we compare \myName\ with four state-of-the-art techniques: DiWi~\cite{DiWi}, RecBi~\cite{RecBi}, ODFL~\cite{ODFL}, and ETEM~\cite{ETEM}. DiWi, RecBi, and ETEM isolate bugs by mutating the failing test to generate witness programs that trigger or avoid the bug, and compute suspiciousness from the coverage of those tests. They differ in mutation and search strategies. ODFL instead explores fine-grained optimization configurations and applies SBFL to rank suspicious elements.

Unlike these input-driven techniques, \myName\ performs bug isolation by analyzing the compilation process. \myName\ divides the compilation process into individually executable steps, identifies bug-causing steps through causal analysis, and ranks suspicious compiler elements by analyzing each step's corresponding compiler source code.

DiWi, RecBi, and ETEM generate mutated witnesses in an open-ended loop with a time limit, making their performance time-sensitive. For fairness, we evaluate them under:

\begin{itemize}
    \item \textbf{\textnormal{X}}: 
    Uses their original paper setting (1 hour per bug).
    \item \textbf{$\textnormal{X}_\textnormal{limT}$}: 
    A time-constrained variant that uses the same per-bug runtime as \myName.
\end{itemize}
where \textbf{\textnormal{X}} refers to DiWi, RecBi, or ETEM. 

Additionally, we repeated each experiment five times and reported the average results to mitigate the impact of randomness in DiWi, RecBi, and ETEM (ODFL is deterministic).

To answer RQ2, we compare \myName\ with ablated variants to assess the impact of each key component:

\begin{itemize}
    \item \textbf{$\textnormal{\myName}_\textnormal{noDel}$}: This variant does not perform any deletion of the step sequence; instead, it directly identifies steps on the original sequence and collects the coverage difference before and after deleting steps in the original sequence.
    \item \textbf{$\textnormal{\myName}_\textnormal{rand}$}: This variant does not adopt the backward tail prune strategy, but instead applies a random order steps identification and deletion.
    \item \textbf{$\textnormal{\myName}_\textnormal{mbfl}$}: This variant does not use our proposed suspiciousness distribution formula; instead, it adopts traditional Mutation-Based Fault Localization (MBFL): by treating the deletion of a step as a mutation of the statements covered by that step, then computing suspiciousness using the MBFL formula Metallaxis~\cite{Metallaxis}.
    \item \textbf{$\textnormal{\myName}_\textnormal{sbfl}$}: This variant does not use our proposed suspiciousness distribution formula; instead, it adopts traditional Spectrum-Based Fault Localization (SBFL) Ochiai~\cite{SBFL-Ochiai} by treating the failing execution before deleting a step as failing test cases, and the passing execution after deleting a step as passing test cases.  
\end{itemize}
By comparing \myName\ with its two variants \textnormal{noDel} and \textnormal{rand}, we investigate the effectiveness of our tail step pruning strategy. By comparing \myName\ with the two variants \textnormal{mbfl} and \textnormal{sbfl}, we examine the effectiveness of our proposed suspiciousness distribution formula.

\subsection{Evaluation Metrics}

In compiler bug isolation, the output is a ranked list of program elements, where elements ranked higher are considered more likely to be the root cause of the bug. To evaluate the effectiveness of each technique, we adopt the following widely used~\cite{DiWi, RecBi, ODFL, ETEM, 10.1145/2931037.2931049, Moon2014AskTM, 10.1145/3092703.3092717} metrics:

\begin{itemize}
    \item Top-$n$: The number of bugs for which at least one buggy program element appears within the top-$n$ positions of the ranked list (with $n \in \{1, 3, 5, 10\}$). Higher values indicate better bug isolation effectiveness.
    \item Mean First Rank (MFR) / Mean Average Rank (MAR): The average rank of the first/all buggy program element in the result list across all bugs. Lower values indicate better bug isolation effectiveness.
    \item Run-Time: The total time required by each bug isolation technique to complete analysis for a given bug.
\end{itemize}

We conduct experiments at both the file level and the function level. Accordingly, the program element in the above metrics refers to either a source file or a function.

\subsection{Implementation}

\myName\ is implemented in Python and uses Gcov~\cite{GCOV} 11.4.0 to collect line-level coverage information. We used publicly available implementations of DiWi, RecBi, ETEM, and ODFL. 

The LLVM experiments were conducted in a Docker container on a workstation with 2×Intel Xeon Gold 5218R CPUs, 250GB RAM, and 4×NVIDIA RTX 3090 GPUs (24GB, CUDA 12.2)\footnote{\myName\ does not use GPUs, but RecBi and ETEM do, so we report the GPU setup.}, running Ubuntu 22.04.3 LTS. The GCC experiments were conducted on a separate server equipped with 2×Intel Xeon Platinum 8468V CPUs (192 hardware threads, 3.80GHz), 512GB RAM, and 3×NVIDIA H200 NVL GPUs (144GB, CUDA 12.4), running Ubuntu 22.04.5 LTS.

%% file: CompSCAN/5-results_and_analysis.tex
\section{Results And Analysis}\label{sec:results_analysis}



\subsection{RQ1: Effectiveness}\label{sec:results_analysis:RQ1}

Table~\ref{tab:Effectiveness-llvm} presents a comparison between \myName\ and existing techniques in terms of bug isolation effectiveness on the 125-bug LLVM dataset. The first column lists the names of the techniques, and the subsequent columns report the results of each isolation metric, including Top-1/3/5/10, MFR, and MAR.

\begin{table*}[t]
\footnotesize
\centering
\caption{Bug Isolation Effectiveness on LLVM}
\begin{tabular}{l|rr|rr|rr|rr|rr|rr}
\hline
Approach & Top1 & $\Uparrow_{Top1}$ & Top3  &  $\Uparrow_{Top3}$ & Top5 & $\Uparrow_{Top5}$ & Top10 & $\Uparrow_{Top10}$ & MFR & $\Uparrow_{MFR}$ & MAR & $\Uparrow_{MAR}$ \\ 

\hline
\rowcolor{gray!20} \myName 
& 32 & - & 53 & - & 62 & - & 78 & - & 20.80 & - & 24.06 & - \\ 
\hline
\hline
ODFL   
& 24 & 33.33 & 35 & 51.43 & 41 & 51.22 & 59 & 32.20 & 31.65 & (-)34.28 & 33.57 & (-)28.33 \\
ETEM    
& 18.20 & 75.82 & 25.80 & 105.43 & 36.00 & 72.22 & 55.40 & 40.79 & 27.98 & (-)25.66 & 38.63 & (-)37.72\\
RecBi   
& 9.40 & 240.43 & 17.00 & 211.76 & 25.60 & 142.19 & 41.80 & 86.60 & 46.42 & (-)55.19 & 58.91 & (-)59.16 \\
DiWi    
& 8.20 & 290.24 & 15.60 & 239.74 & 23.40 & 164.96 & 42.20 & 84.83 & 37.45 & (-)44.46 & 47.60 & (-)49.45 \\
\hline
$\textnormal{ETEM}_\textnormal{limT}$
&  12.40 & 158.06 & 23.60 & 124.58 & 34.20 & 81.29 & 57.60 & 35.42 & 36.60 & (-)43.17 & 45.04 & (-)46.58\\ 
$\textnormal{RecBi}_\textnormal{limT}$
& 8.20 & 290.24 & 15.60 & 239.74 & 24.40 & 154.10 & 44.20 & 76.47 & 36.25 & (-)42.62 & 46.49 & (-)48.25 \\ 
$\textnormal{DiWi}_\textnormal{limT}$
& 7.40 & 332.43 & 18.80 & 181.91 & 24.80 & 150.00 & 43.40 & 79.72 & 45.30 & (-)54.08 & 57.95 & (-)58.48\\ 
\hline
\hline
$\textnormal{\myName}_\textnormal{noDel}$
& 17 & 88.24 & 32 & 65.62 & 42 & 47.62 & 50 & 56.00 & 85.5 & (-)75.67 & 93.98 & (-)74.40 \\ 
$\textnormal{\myName}_\textnormal{rand}$
& 24 & 33.33 & 42 & 26.19 & 55 & 12.73 & 65 & 20.00 & 57.83 & (-)64.03 & 66.66 & (-)63.91\\ 
\hline
$\textnormal{\myName}_\textnormal{mbfl}$
& 3 & 966.67 & 7 & 657.14 & 10 & 520.00 & 14 & 457.14 & 310.86 & (-)93.31 & 326.74 & (-)92.64 \\ 
$\textnormal{\myName}_\textnormal{sbfl}$
& 27 & 18.52 & 47 & 12.77 & 59 & 5.08 & 70 & 11.43 & 100.94 & (-)79.39 & 107.74 & (-)77.67 \\ 
\hline
\end{tabular}

\begin{minipage}{\textwidth}
\vspace{1mm}
``$\Uparrow_{*}$'' refers to the improvement rate of \myName\ over a compared approach in terms of the metric ``$*$''.
\end{minipage}
\label{tab:Effectiveness-llvm} 
\end{table*}

\subsubsection{Bug Isolation Effectiveness}
Overall, as shown in Table~\ref{tab:Effectiveness-llvm}, for all 125 bugs, \myName\ achieves Top-1/3/5/10 scores of 32, 53, 62, and 78, respectively. In other words, the root cause buggy file for 32, 53, 62, and 78 bugs appears within the top 1, 3, 5, and 10 positions in \myName’s output ranking list, respectively. The MFR and MAR scores are 20.80 and 24.06, indicating that the average rank of the most suspicious buggy file per bug is 20.80, and the average rank of all buggy files across bugs is 24.06.

\myName\ outperforms existing techniques in compiler bug isolation. Specifically, as shown in Table~\ref{tab:Effectiveness-llvm}, \myName\ achieves improvements over the state-of-the-art mutation-based technique ETEM by 75.82\%, 105.43\%, 72.22\%, and 40.79\% in Top-1/3/5/10, and by 25.66\% and 37.72\% in MFR and MAR, respectively. 
Compared to the state-of-the-art optimization-search-based technique ODFL, \myName\ shows improvements of 33.33\%, 51.43\%, 51.22\%, and 32.20\% in Top-1/3/5/10, and 34.28\% and 28.33\% in MFR and MAR, respectively. 
For other compiler bug isolation techniques, \myName\ achieves gains over RecBi of 240.43\%, 211.76\%, 142.19\%, and 86.60\% in Top-1/3/5/10, and 55.19\% and 59.16\% in MFR and MAR, respectively. Compared to DiWi, \myName\ improves performance by 290.24\%, 239.74\%, 164.96\%, and 84.83\% in Top-1/3/5/10, and by 44.46\% and 49.45\% in MFR and MAR.

These results demonstrate that \myName\ achieves superior compiler bug isolation performance compared to existing techniques, with notable improvements on the Top-n/MFR/MAR metrics.

\begin{figure}[htbp]
\centering
\subfigure[Top-1 Bugs Intersection]{
    \includegraphics[width=0.38\textwidth]{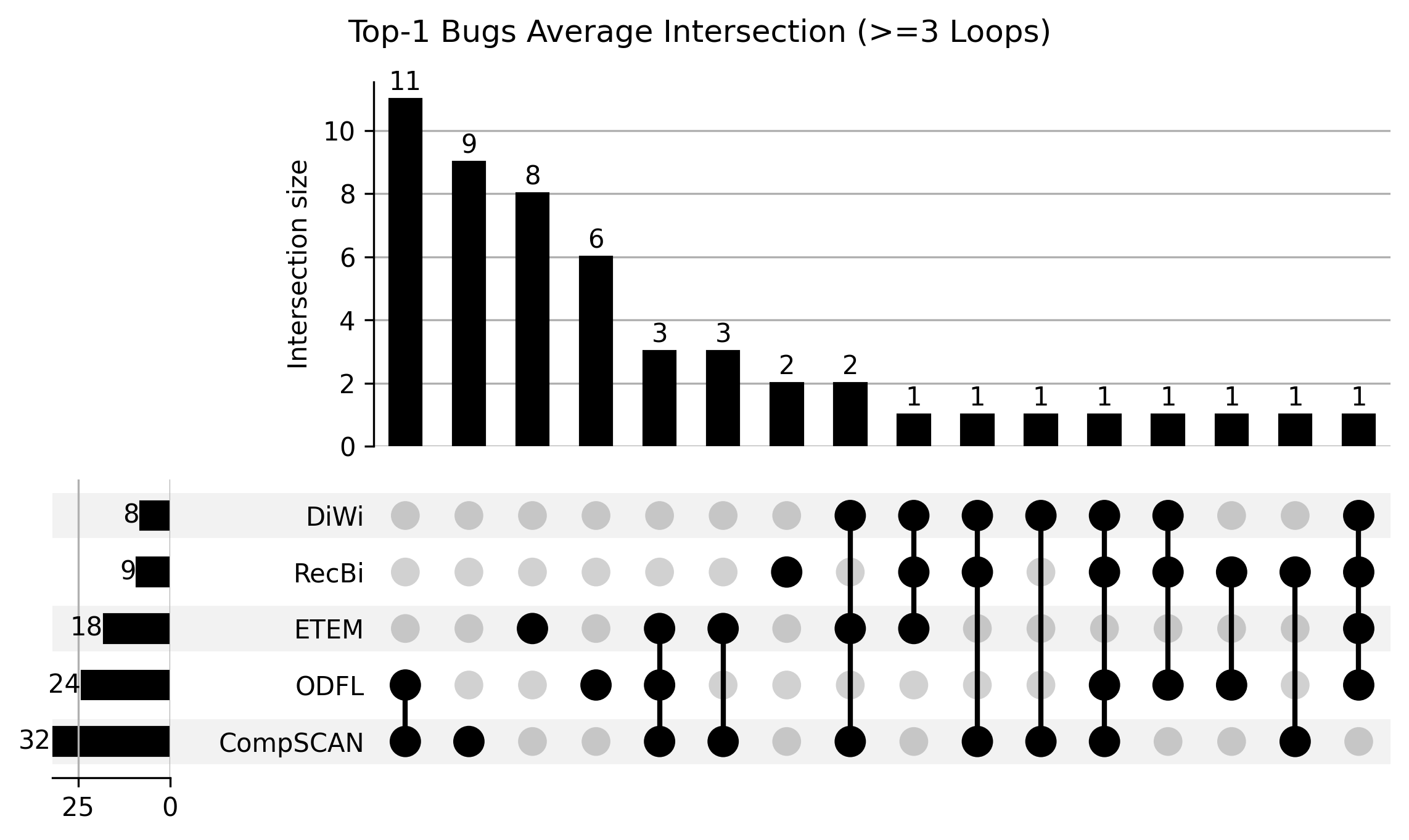}
    \label{fig:top1-intersection}
}
\subfigure[Top-5 Bugs Intersection]{
    \includegraphics[width=0.48\textwidth]{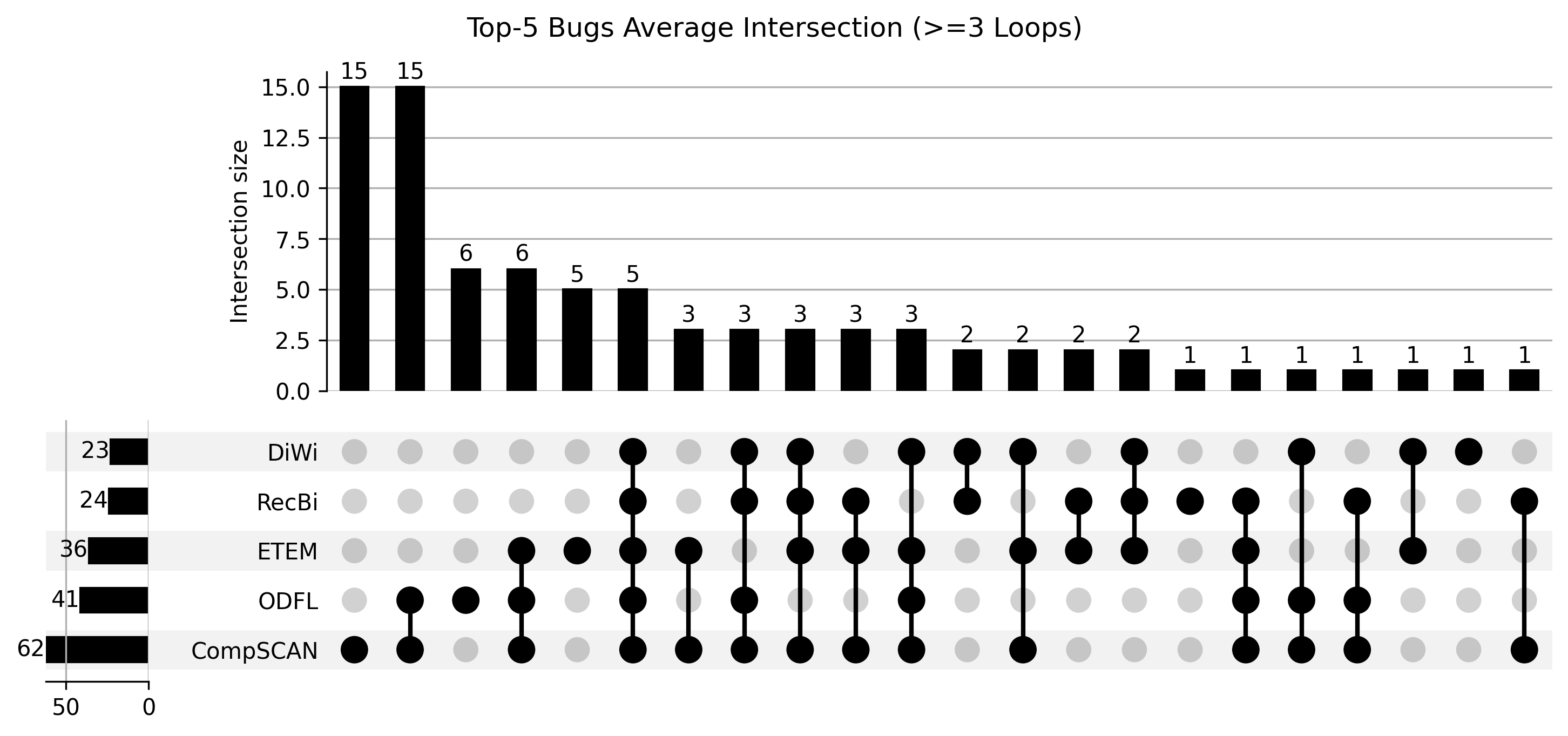}
    \label{fig:top5-intersection}
}
\caption{Intersection of Top-1 and Top-5 Bugs}
\Description{Visualization of the intersections of bugs successfully isolated at Top-1 and Top-5 by different techniques. Each horizontal bar represents the total number of bugs isolated by a technique, while the matrix of black dots indicates combinations of techniques.}
\label{fig:topk-intersection}
\end{figure}

Figure~\ref{fig:topk-intersection} illustrates the intersections of bugs successfully isolated at Top-1 and Top-5 by different techniques. Each horizontal bar shows the total number of bugs a technique successfully isolates. Next to these horizontal bars is a matrix of black dots, where each column of filled dots represents a specific combination of techniques. A dot is filled if the corresponding technique is included in that combination. The vertical bar above each dot column indicates the number of bugs that are successfully isolated (within Top-1 or Top-5) by exactly the combination of techniques represented by the black dots in that column. 
For techniques such as ETEM, RecBi, and DiWi, which we repeated five times due to inherent randomness, we report the set of bugs isolated at Top-1/5 by requiring that a bug’s rank be $\leq 1$ (or $\leq 5$) in at least three out of five runs. This majority-vote criterion ensures robustness against variance across repeated executions.

For example, the second tallest bar in Figure~\ref{fig:top1-intersection}, aligned with a column that has only the dot for \myName\ filled, shows that 9 bugs are uniquely isolated at Top-1 by \myName\ alone; the first tallest bar in Figure~\ref{fig:top1-intersection}, aligned with a column that has filled dots for \myName\ and ODFL, shows that 11 bugs are uniquely isolated at Top-1 by \myName\ and ODFL, and none of the other techniques can isolate these 3 bugs at Top-1. 

Similarly, as shown in Figure~\ref{fig:top5-intersection}, \myName\ isolates the highest number (i.e., 15) of unique bugs at Top-5.
Therefore, the results in Figure~\ref{fig:top1-intersection} and Figure~\ref{fig:top5-intersection} indicate that \myName\ successfully isolates the largest number of unique bugs within the Top-1 and Top-5 ranking compared to other techniques. This demonstrates the distinctive effectiveness of \myName\ in isolating compiler bugs that are otherwise challenging for existing techniques.

\subsubsection{Runtime Evaluation}

\begin{table}[t]
\caption{Running Time Comparison on LLVM}
\begin{center}
\begin{tabular}{c|cc|cc|cc}
\hline
\multirow{2}{*}{Approach} & \multicolumn{6}{c}{Running Time (s)} \\
\cline{2-7}
 & Avg & $\Uparrow_{Avg}$ & Min & $\Uparrow_{Min}$ & Max & $\Uparrow_{Max}$ \\
\hline
\rowcolor{gray!20} \myName 
& 168.16 & - & 34.97 & - & 941.46 & - \\
\hline
ODFL 
& 523.83 & (-)67.90 & 209.25 & (-)83.29 & 1206.07 & (-)21.94 \\
DiWi/RecBi/ETEM 
& 3600.00*5 & (-)99.07 & 3600.00*5 & (-)99.81 & 3600.00*5 & (-)94.77 \\
\hline
\end{tabular}
\end{center}
\label{tab:RQ1-Search-Time}
\end{table}

Compared to prior techniques, \myName\ requires significantly less time to isolate bugs. Table~\ref{tab:RQ1-Search-Time} reports the average (Avg), minimum (Min), and maximum (Max) time each technique takes to isolate a single bug. Notably, RecBi, DiWi, and ETEM operate under predefined time budgets, and their runtimes are determined by these preset limits. In Table~\ref{tab:RQ1-Search-Time}, we adopt the default configurations as specified in their original studies. 

As shown in Table~\ref{tab:RQ1-Search-Time}, \myName\ achieves the shortest isolating time (i.e., an average of 168.16s per bug), demonstrating the efficiency. We further analyzed the reason why \myName\ is faster than ODFL. First, our technique focuses on the executed compilation steps, whereas ODFL broadly considers all fine-grained optimization options, regardless of whether this option is executed or not. 
Second, ODFL depends on enumerating a large number of fine-grained optimization configurations for statistical analysis, while \myName, by incorporating the tail-prune strategy, only needs to traverse the compilation step sequence once in a backward manner. Consequently, the number of compilations required by \myName\ is smaller than that of ODFL. 
In summary, \myName\ is designed carefully with computational efficiency in mind, and the efficient algorithm leads to lower bug isolation time.

To further ensure fair comparison with ETEM, RecBi, and DiWi, we also conducted additional experiments where the execution time of ETEM, RecBi, and DiWi was limited to match that of \myName, comparing the bug isolation effectiveness under equal time constraints. 
As shown in Table~\ref{tab:Effectiveness-llvm}, the rows labeled with $\textnormal{X}_\textnormal{limT}$ ($\textnormal{X}\in\{\textnormal{ETEM, RecBi, DiWi}\}$) correspond to the performance of each baseline when restricted to the same runtime as \myName. The results indicate that \myName\ outperforms across all metrics—including Top-n, MFR, and MAR—under identical time budgets, further demonstrating \myName's efficiency and effectiveness.

\subsubsection{Statistical Test}
Following prior work~\cite{ODFL, Tu2024LLM4CBI, U-test-p-0.05}, we conducted a Mann–Whitney U-test with a significance level of 0.05 on the 125 LLVM bugs to compare \myName\ against state-of-the-art approaches, and we measured the effect size using Vargha and Delaney’s $\hat{A}_{12}$ statistic. Consistent with prior studies, we utilize the R programming language to perform the test, obtaining the $p$-values and $\hat{A}_{12}$ values.  

Here, the effect size $\hat{A}_{12}$ is computed as:  
$$
\hat{A}_{12} = \frac{(R_1 / m) - (m+1)/2}{n},
$$
where $R_1$ denotes the rank sum of the first group under comparison, $m$ is the number of observations in the first group, and $n$ is the number of observations in the second group.  

In this context, a Mann–Whitney U-test with a $p$-value less than 0.05 indicates that the performance difference between \myName\ and the compared technique is statistically significant. Moreover, the $\hat{A}_{12}$ statistic measures the probability that approach A produces higher values than approach B, with $\hat{A}_{12} \in [0,1]$. A value of $\hat{A}_{12} = 0.5$ suggests no difference between the two approaches, while larger values indicate stronger evidence that A outperforms B.

\begin{table}[t]
\caption{Statistical Test on LLVM}
\centering
\begin{minipage}{0.48\linewidth}
\centering
\begin{tabular}{c|c|cc}
\hline
Comparison & Metric & p-value & $\hat{A}_{12}$ \\
\hline
\multirow{6}{*}{\myName\ vs. ODFL} 
& Top1 & 0.0040 & 1.0000 \\
& Top3 & 0.0040 & 1.0000 \\
& Top5 & 0.0040 & 1.0000 \\
& Top10 & 0.0040 & 1.0000 \\
& MFR & 0.0040 & 1.0000 \\
& MAR & 0.0040 & 1.0000 \\
\hline
\multirow{6}{*}{\myName\ vs. ETEM} 
& Top1 & 0.0071 & 1.0000 \\
& Top3 & 0.0075 & 1.0000 \\
& Top5 & 0.0075 & 1.0000 \\
& Top10 & 0.0073 & 1.0000 \\
& MFR & 0.0075 & 1.0000 \\
& MAR & 0.0075 & 1.0000 \\
\hline
\end{tabular}
\end{minipage}
\hfill
\begin{minipage}{0.48\linewidth}
\centering
\begin{tabular}{c|c|cc}
\hline
Comparison & Metric & p-value & $\hat{A}_{12}$ \\
\hline
\multirow{6}{*}{\myName\ vs. RecBi} 
& Top1 & 0.0073 & 1.0000 \\
& Top3 & 0.0073 & 1.0000 \\
& Top5 & 0.0071 & 1.0000 \\
& Top10 & 0.0075 & 1.0000 \\
& MFR &  0.0075 & 1.0000 \\
& MAR &  0.0075 & 1.0000 \\
\hline
\multirow{6}{*}{\myName\ vs. DiWi} 
& Top1 & 0.0071 & 1.0000 \\
& Top3 & 0.0073 & 1.0000 \\
& Top5 & 0.0073 & 1.0000 \\
& Top10 & 0.0073 & 1.0000 \\
& MFR & 0.0075 & 1.0000 \\
& MAR & 0.0075 & 1.0000 \\
\hline
\end{tabular}
\end{minipage}
\label{tab:RQ1-Statistic}
\end{table}

The statistical results are summarized in Table~\ref{tab:RQ1-Statistic}. As shown in the table, for all metrics, the $p$-values of \myName\ compared with any baseline are below 0.05, demonstrating that the differences are statistically significant.

Moreover, for all techniques, the effect sizes $\hat{A}_{12}$ are greater than 0.999, indicating that \myName\ has a high probability of achieving better results than the state-of-the-art methods.

In summary, the statistical tests provide strong evidence that \myName\ significantly outperforms existing techniques.

\subsection{RQ2: Ablation Study}\label{sec:results_analysis:RQ2}


The last four rows of Table~\ref{tab:Effectiveness-llvm} report the results of our ablation study.

Compared to the variant $\textnormal{\myName}_\textnormal{noDel}$, which disables the tail pruning strategy, \myName\ achieves improvements across all metrics. At the file level, Top-1/3/5/10 accuracy increased by 88.24\%, 65.62\%, 47.62\%, and 56.00\%, respectively, while MFR and MAR improved by 75.67\% and 74.40\%.  
Similarly, compared to the variant $\textnormal{\myName}_\textnormal{rand}$, which performs step identification and deletion in a random order instead of following the backward tail pruning strategy, \myName\ shows consistent improvements across all metrics. At the file level, Top-1/3/5/10 accuracy increased by 33.33\%, 26.19\%, 12.73\%, and 20.00\%, respectively, while MFR and MAR improved by 64.03\% and 63.91\%.

These results highlight the effectiveness of the tail-pruning strategy. Intuitively, only by deleting steps in reverse order can we ensure that for each bug-causing step, the collected coverage difference minimally includes irrelevant downstream statements. Not deleting steps or in a random order would likely leave additional unaffected statements in the coverage difference, reducing the precision of bug localization.

Compared to the variant $\textnormal{\myName}_\textnormal{mbfl}$, which employs the traditional MBFL formula Metallaxis for suspiciousness calculation, \myName\ instead adopts our formula with the $\tfrac{1}{|\text{diff}|}$ factor and achieves substantial improvements across all metrics. At the file level, Top-1/3/5/10 accuracy increased by 966.67\%, 657.14\%, 520.00\%, and 457.14\%, respectively, while MFR and MAR improved by 93.31\% and 92.64\%.  
These results demonstrate the effectiveness of introducing the $\tfrac{1}{|\text{diff}|}$ factor in our suspiciousness computation. Without this factor, the traditional MBFL approach fails to account for the scenario where a single mutant may involve a large number of statements: regardless of the size of the coverage difference, all statements would otherwise receive the same suspiciousness score. This leads to severe bias in ranking, whereas our formulation naturally distributes suspiciousness in proportion to the coverage difference size, producing far more accurate bug isolation.

Compared to the variant $\textnormal{\myName}_\textnormal{sbfl}$, which applies the traditional SBFL formula Ochiai for suspiciousness computation, \myName\ achieves consistent improvements across all metrics. At the file level, Top-1/3/5/10 accuracy increased by 18.52\%, 12.77\%, 5.08\%, and 11.43\%, respectively, while MFR and MAR improved by 79.39\% and 77.67\%.  
These results validate the effectiveness of our coverage-difference–based suspiciousness formulation. In our setting, where bug-causing steps can be accurately identified, the number of such steps is inherently small. Consequently, statistical techniques like ODFL or SBFL lack sufficient data for reliable inference. Nevertheless, even $\textnormal{\myName}_\textnormal{sbfl}$ already outperforms prior state-of-the-art methods, highlighting the intrinsic advantage of our step-driven design.

In summary, each component of \myName\ plays an important role in the final bug isolation performance.

\subsection{RQ3: Generalizablility}

\begin{table*}[t]
\footnotesize
\centering
\caption{Bug Isolation Effectiveness on GCC}
\begin{tabular}{l|rr|rr|rr|rr|rr|rr}
\hline
Approach & Top1 & $\Uparrow_{Top1}$ & Top3  &  $\Uparrow_{Top3}$ & Top5 & $\Uparrow_{Top5}$ & Top10 & $\Uparrow_{Top10}$ & MFR & $\Uparrow_{MFR}$ & MAR & $\Uparrow_{MAR}$ \\ 

\hline
\rowcolor{gray!20} \myName
& 18 & - & 32 & - & 38 & - & 45 & - & 5.36 & - & 5.37 & - \\ 
\hline
\hline
ODFL
& 14 & 28.57 & 22 & 45.45 & 28 & 35.71 & 42 & 7.14 & 16.53 & (-)67.57 & 18.96 & (-)71.68 \\
ETEM    
& 16.40 & 9.76 & 30.80 & 3.90 & 37.40 & 1.60 & 43.40 & 3.69 & 8.01 & (-)33.08 & 8.61 & (-)37.63\\
RecBi   
& 10.00 & 80.00 & 24.40 & 31.15 & 29.40 & 29.25 & 35.60 & 26.40 & 14.95 & (-)64.15 & 15.64 & (-)65.66 \\
DiWi   
& 8.20 & 119.51 & 19.00 & 68.42 & 24.60 & 54.47 & 34.00 & 32.35 & 15.41 & (-)65.22 & 16.12 & (-)66.69 \\
\hline
$\textnormal{ETEM}_\textnormal{limT}$
& 14.60 & 23.29 & 28.20 & 13.48 & 34.40 & 10.47 & 43.80 & 2.74 & 7.53 & (-)28.82 & 8.33 & (-)35.53 \\ 
$\textnormal{RecBi}_\textnormal{limT}$
& 6.60 & 172.73 & 12.20 & 162.30 & 18.00 & 111.11 & 27.20 & 65.44 & 24.01 & (-)77.68 & 25.70 & (-)79.11 \\
$\textnormal{DiWi}_\textnormal{limT}$
& 6.00 & 200.00 & 12.80 & 150.00 & 16.80 & 126.19 & 27.00 & 66.67 & 24.91 & (-)78.48 & 27.05 & (-)80.15 \\
\hline
\end{tabular}
\begin{minipage}{\textwidth}
\vspace{1mm}
``$\Uparrow_{*}$'' refers to the improvement rate of \myName\ over a compared approach in terms of the metric ``$*$, for MFR/MAR a lower value is better''.
\end{minipage}
\label{tab:Effectiveness-gcc} 
\end{table*}

\begin{table}[t]
\caption{Running Time Comparison on GCC}
\begin{center}
\begin{tabular}{c|cc|cc|cc}
\hline
\multirow{2}{*}{Approach} & \multicolumn{6}{c}{Running Time (s)} \\
\cline{2-7}
 & Avg & $\Uparrow_{Avg}$ & Min & $\Uparrow_{Min}$ & Max & $\Uparrow_{Max}$ \\
\hline
\myName 
& 76.01 & - & 7.75 & - & 303.47 & - \\
\hline
ODFL 
& 130.58 & (-)41.79 & 27.94 & (-)72.26 & 646.37 & (-)53.05 \\
DiWi/RecBi/ETEM 
& 3600.00*5 & (-)99.58 & 3600.00*5 & (-)99.96 & 3600.00*5 & (-)98.31 \\
\hline
\end{tabular}
\end{center}
\label{tab:Run-Time-GCC}
\end{table}

To evaluate the generality of \myName, we conducted experiments not only on LLVM but also on another widely used compiler, GCC, to compare the effectiveness of our technique against existing approaches. Unlike LLVM, obtaining the execution order of GCC's mid-end optimizations is challenging: GCC does not provide an option similar to LLVM's \texttt{--print-pipeline-passes} that exposes compilation steps and their sequence. Moreover, GCC optimizations may overlap. For example, in GCC bug \#58068 (compiler version r201397), the \texttt{-fdce} optimization consists of two sub-passes, \texttt{rtl-ud\_dce} and \texttt{rtl-rtl\_dc}, while \texttt{-fdse} contains \texttt{rtl-dse1} and \texttt{rtl-dse2}. The actual execution order of these four sub-passes is \texttt{rtl-dse1}, \texttt{rtl-ud\_dce}, \texttt{rtl-dse2}, \texttt{rtl-rtl\_dc}, meaning that \texttt{-fdce} and \texttt{-fdse} partially overlap in execution.

Nevertheless, \myName\ demonstrates generality and can operate even when the precise execution order of compilation steps is unavailable. Consequently, we also conducted comparative experiments on GCC. For this evaluation, we chose the last sub-pass of each GCC optimization as the representative order for that optimization step\footnote{The reason for selecting the last sub-pass of each optimization as the ordering basis is that the tail-pruning strategy aims to minimize the influence of subsequent steps when collecting statements corresponding to each step. By using the last sub-pass to represent the optimization, we ensure that all deletable sub-passes following it have already been processed. If any deletable sub-pass were still present, the corresponding original optimization step would have been handled and removed earlier.}, while keeping all other settings unchanged. We then compared the effectiveness of \myName\ against existing techniques on a dataset of 60 real GCC bugs.  

Table~\ref{tab:Effectiveness-gcc} presents the effectiveness of various techniques on GCC. As shown, \myName\ continues to outperform existing approaches. For all 60 bugs, \myName\ achieves Top-1/3/5/10/MFR/MAR scores of 18, 32, 38, 45, 5.36, and 5.37, respectively, representing improvements over ETEM by 9.76\%, 3.90\%, 1.60\%, 3.69\%, 33.08\%, and 37.63\%, and over ODFL by 28.57\%, 45.45\%, 35.71\%, 7.14\%, 67.57\%, and 71.68\%, respectively. Table~\ref{tab:Run-Time-GCC} reports the runtime required by each technique to isolate bugs on GCC. It can be observed that \myName\ consistently requires the least average runtime, further demonstrating its efficiency.  

These results indicate that even in scenarios where obtaining the precise execution order of compilation steps is difficult, \myName\ still outperforms state-of-the-art techniques in bug isolation. At the same time, the runtime remains lower than competing methods, highlighting the generality of \myName. This also underscores that, compared to existing compiler bug isolation techniques that rely on generating a large number of tests for statistical analysis, \myName's strategy that accurately identifies bug-causing steps and the corresponding statements, and distributing suspiciousness based on the set of affected statements, yields both higher isolation effectiveness and greater efficiency.

%% file: CompSCAN/6-discussion.tex
\section{Threats to Validity}\label{sec:discussion}

Our study involves several threats to validity, which we have mitigated to the best extent possible.

Internal threats come from the implementations of \myName\ and baseline techniques. To mitigate this, we utilized the latest open-source, reproducible packages from DiWi, RecBi, ETEM, and ODFL. Furthermore, our \myName\ was implemented with Python standard libraries and thoroughly reviewed by the authors. 

External threats lie in the choice of the dataset. To mitigate this, we used all the LLVM and GCC bugs from prior studies and added 30 recent bugs to increase dataset diversity. 

Construct threats concern evaluation metrics. Following prior work, we used widely accepted Top-n, MFR, MAR, and runtime metrics. Additionally, for techniques involving randomness (DiWi, RecBi, and ETEM), we averaged results over five runs to reduce the impact of randomness.


%% file: CompSCAN/7-related_work.tex
\section{Related Work}\label{sec:related_work}

\subsection{Automated Fault Localization}
Automated fault localization aims to identify the locations of faults within a program. 
A wide range of fault localization techniques have been proposed~\cite{wong2016survey}, including earlier approaches such as Spectrum-Based Fault Localization (SBFL)~\cite{SBFL-Ochiai, SBFL-Ochiai2, SBFL-tarantula, SBFL-Dstar}, Mutation-Based Fault Localization~\cite{MUSE, Metallaxis, Moon2014AskTM}, and slice-based approaches~\cite{mao2014slice, wang2014slice}, as well as more recent learning-based approaches and so on~\cite{wong2011learning, xuan2014learning, li2021learning, zeng2022probabilistic, chen2024learning}.
However, most existing techniques are designed for general software and assume test cases and program behaviors that do not typically reflect the characteristics of compiler internals. Recent studies~\cite{DiWi, RecBi} have shown that these techniques are not directly applicable to compiler bug isolation due to the large codebase and complexity of compiler pipelines.

\myName\ inherits ideas from MBFL for computing suspiciousness scores. MBFL relies on creating program mutants and observing their impact on the test outcomes. Each mutant is used to estimate how likely a statement is to be faulty: statements whose mutations frequently change the results of failing tests are considered more suspicious~\cite{MUSE, Metallaxis}. 
Unlike traditional MBFL, which operates at the level of individual statements in application programs, \myName\ treats the removal of a bug-causing compilation step as a “mutation” and computes suspiciousness over the associated coverage differences, enabling effective bug isolation in the compiler setting.

\subsection{Compiler Bug Isolation}
As general fault localization techniques are not well-suited for compilers, specialized approaches have emerged, which can be broadly categorized into two types: witness-test-program–based and fine-grained optimization–based methods.

Witness-test-program–based techniques, such as DiWi~\cite{DiWi}, RecBi~\cite{RecBi}, LLM4CBI~\cite{Tu2024LLM4CBI}, and ETEM~\cite{ETEM}, generate test programs, called witness programs, that either trigger or avoid bugs, thereby helping to reduce suspicion of non-faulty compiler components. These methods typically adopt search strategies to effectively explore the vast mutation space. DiWi employs a Metropolis-Hastings algorithm~\cite{MCMC-MH} to guide mutations over variables, operators, and constants. RecBi expands mutation coverage to control-flow constructs and utilizes reinforcement learning~\cite{reinforcement-learning, RL-ZTE-2} for search. LLM4CBI leverages similar mutation strategies but delegates test generation to large language models via tailored prompts. ETEM further introduces compiler-specific feature mutations and performs test program search using joint reinforcement learning.

Fine-grained optimization–based techniques, such as LocSeq~\cite{Zhou2022LocSeq} and ODFL~\cite{ODFL}, isolate optimization bugs by exploring optimization configurations rather than test programs. LocSeq utilizes a constrained genetic algorithm guided by the fitness function, while ODFL categorizes options into bug-related and bug-free sets and incrementally enumerates configurations.

Unlike prior techniques that either mutate programs or explore optimization flags, \myName analyzes the compiler's internal process directly. By applying analysis over the sequence of compilation steps, \myName identifies and ranks suspicious code elements within the compiler.

Techniques targeting specialized compilers, such as Java JIT compilers, have been proposed to localize optimization and code generation bugs by comparing instruction-level traces coverage of passing and failing tests~\cite{lim2021JIT,lim2023JITbackend}.
These approaches rely on fine-grained execution traces, which are only feasible in JIT compilers due to JIT's small codebase and simplified, specialized compilation pipelines. However, general-purpose compilers such as GCC and LLVM fundamentally differ from JIT compilers in both architecture and scale. They typically consist of large codebases and complex, multi-phase compilation pipelines, making instruction-level trace collection prohibitively expensive and difficult to manage in practice.

%% file: CompSCAN/8-conclusion.tex
\section{Conclusion}\label{sec:conclusion}

We present \myName, a novel compiler bug isolation technique that leverages the structure of the compilation step sequence. \myName\ begins by identifying the initial failing sequence and determining the bug-causing steps. It then collects the corresponding compiler source code and identifies suspicious statements. To reduce noise from irrelevant code, \myName\ applies tail pruning of non-essential steps, minimizing downstream effects. Finally, it ranks the remaining compiler elements by suspiciousness to support efficient and precise bug isolation.  

We evaluated \myName\ on 125 real LLVM bugs and 60 real GCC bugs, comparing it with state-of-the-art techniques and performing ablation studies. The results demonstrate that \myName\ significantly outperforms existing approaches in both effectiveness and efficiency, and that each of its components contributes meaningfully to its overall performance.